\documentclass[applsci,submit,pdftex,moreauthors,12pt,a4paper]{mdpi} 
\firstpage{1} 
\pubvolume{1}
\issuenum{1}
\articlenumber{0}
\pubyear{2025}
\copyrightyear{2025}
\datereceived{December 2025} 
\dateaccepted{2026} 
\datepublished{2026} 
\hreflink{https://doi.org/} 
\doinum{}
\pdfoutput=1 

\usepackage{todonotes}
\usepackage{arydshln}

\Title{Spectrogram features for audio and speech analysis}

\TitleCitation{Spectrogram features for audio and speech analysis}

\Author{Ian McLoughlin $^{1,\dagger}$\orcidA{}*, Lam Pham $^{2}$\orcidB{}, Yan Song$^{3}$\orcidC{}, Xiaoxiao Miao$^{1}$\orcidD{}, Huy Phan$^{4}$\orcidE{}, Pengfei Cai$^{3}$\orcidF{}, Qing Gu$^{3}$\orcidG{}, Jiang Nan$^{3}$\orcidH{}, Haoyu Song$^{1}$\orcidI{}, Donny Soh$^{1}$\orcidJ{}\textbf{}}

\AuthorNames{Ian McLoughlin, Firstname Lastname and Firstname Lastname}

\isAPAStyle{%
       \AuthorCitation{McLoughlin, I., Lastname, F., \& Lastname, F.}
         }{%
        \isChicagoStyle{%
        \AuthorCitation{Lastname, Firstname, Firstname Lastname, and Firstname Lastname.}
        }{
        \AuthorCitation{McLoughlin, I.; Lastname, F.; Lastname, F.}
        }
}

\address{%
$^{1}$ \quad Singapore Institute of Technology; \{ian.mcloughlin@, 2303822@sit., donny.soh@\}singaporetech.edu.sg\\
$^{2}$ \quad Austrian Institute of Technology; lam.pham@ait.ac.at\\
$^{3}$ \quad The University of Science and Technology of China; \{songy@, cqi525@mail., jiang\_nan@mail., qinggu6@mail.\}ustc.edu.cn\\
$^{4}$ \quad Meta Inc., Reality Labs; huy.phan@ieee.org
}

\corres{Correspondence: ian.mcloughlin@singaporetech.edu.sg}

\firstnote{1 Punggol Coast Road, Singapore 828608}  

\abstract{Spectrogram-based representations have grown to dominate the feature space for deep learning audio analysis systems, and are often adopted for speech analysis also. Initially, the primary motivator for spectrogram-based representations was their ability to present sound as a two dimensional signal in the time-frequency plane, which not only provides an interpretable physical basis for analysing sound, but also unlocks the use of a wide range of machine learning techniques such as convolutional neural networks, that had been developed for image processing. 
A spectrogram is a matrix characterised by the resolution and span of its two dimensions, as well as by the representation and scaling of each element. Many possibilities for these three characteristics have been explored by researchers across numerous application areas, with different settings showing affinity for various tasks.
This paper reviews the use of spectrogram-based representations and surveys the state-of-the-art to question how front-end feature representation choice allies with back-end classifier architecture for different tasks.}

\keyword{Spectrogram; Spectrogram Image Feature, Mel-frequency Spectrogram; Mel Frequency Cepstral Coefficient (MFCC), Constant-Q transform, audio analysis, speech classification} 

\featuredapplication{Spectrogram-based input features have become the most popular choice for deep learning models that classify audio and speech, yet there are many settings related to resolution and representation type. This article surveys those choices and discusses their suitability for different application areas.}
\def\nolinenumbers{} 
\begin{document}


\section{Introduction}

The spectrogram, considered to have been invented at Bell Labs in the 1940s~\citep{koenig_sound_1946} was initially generated by a sound spectrograph machine as a stylus-on-paper plot to visualise the distribution of sound energy in a time-frequency plane. Then, and now, it transforms a one-dimensional sound waveform into a two dimensional image.

Originally popular for ease of visualisation, allowing identification of important structures within a sound signal by eye, the spectrogram became useful in phonetics and various branches of acoustics. Well established by the late 1970s, its information-carrying abilities were highlighted by Victor Zue and Ron Cole who demonstrated that it could be used for speech recognition~\citep{zue1979experiments}.

\begin{figure}[H]
\includegraphics[scale=1]{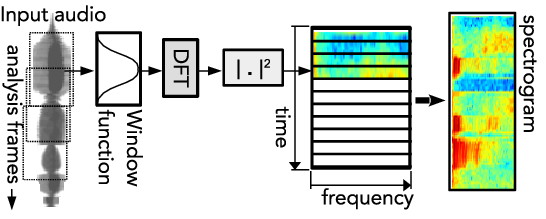}
\caption{Illustration of spectrogram creation from input audio data, as a stack of frequency vectors.\label{fig:spectrogram}}
\end{figure}   
\unskip

\section{Taxonomy of spectrograms}
\label{sec:taxonomy}

At its heart, a spectrogram is nothing more than a two-dimensional picture of sound, usually with one axis representing frequency and the other axis representing time. Individual pixel intensity represents in some way the strength of each frequency element at a particular time instant. In the earliest systems~\citep{koenig_sound_1946} the  frequency axis and intensity were non-linear. Advances in sensors, high quality analogue-to-digital converters and improvements in signal processing led to the ability to form linear spectrograms~\citep{CUPbook2016}. More recently, non-linear representations have again become prevalent, tuned for different tasks, as we will explore in the remainder of this paper.

\subsection{Basic spectrogram}
\label{subsec:basic}

Spectrograms are typically formed as a matrix of stacked frequency vectors, each of which represent the frequency magnitude over a short duration of time, referred to as an analysis frame. 
The frequency magnitude vectors are obtained from an orthogonal time-frequency transform such as a discrete Fourier transform (DFT), fast Fourier transform (FFT) or filterbank (FB). Discrete cosine transform (DCT), modified DCT, discrete wavelet transform (DWT) and many other transforms have been employed~\citep{CUPbook2016}.
Fixed-size analysis frames are typically slices of an input audio waveform of between 10 and 30ms for audible signals. Hence an auditory spectrogram is formed from a stack of frequency vectors obtained from successive frames, as illustrated in Fig.~\ref{fig:spectrogram}.
In almost all cases, to avoid issues with spectral leakage and edge effects (e.g. Gibbs phenomenon~\citep{gibbs}), the slices of input audio are windowed~\citep{ifeachor1993} prior to the time-frequency transform, and because the window functions usually taper to zero or near-zero at the edges of each frame, the frames are overlapped to ensure that frequencies from all regions in the input audio waveform (i.e. including the tapered edge regions), will contribute to the frequency representation~\citep{CUPbook2016}. The overlap between frames is specified either as a percentage (often 50\% overlap), or as a step between frames e.g. 256 sample windows advanced by 128 steps between frames or a 30\,ms window advancing 10\,ms each step. The step can be referred to as an hop, advance or shift. The maximum frequency resolution is limited by the number of samples in the analysis window, and the spectrogram time axis resolution is defined by the step size.

Let us denote an input audio waveform as $x(n)$ and set frame length to be $w_s$ samples for current analysis frame $f$. 
With 50\% overlap between frames, the analysis frame is $x_f(n)=x[f.w_s/2:f.w_s/2+w_s]$. Given a length $w_s$ window function $w(n)$, the spectral magnitude representation $X(k)$ is then,

\begin{equation}
X_f(k) = \left | \sum_{n=0}^{w_s-1} w(n)x_f(n)e^{-{j2{\pi}nk}/{w_s}} \right |     \qquad for~ k=0 \dots w_s-1\\
\label{eqn:eqn1}
\end{equation}

Spectrogram $\mathcal{S}$ is obtained by stacking the frequency vectors directly into a rectangular matrix, i.e. for a time duration of $F$ frames,

\begin{equation}
\label{eq:spectrogramS}
\mathcal{S}_{F,w_s} =
\begin{bmatrix}
X_0 & X_1 & X_2 \dots X_F\\
\end{bmatrix}	
\end{equation}

When used as an input feature to a deep learning system, it is also common that frequency downsampling or pooling happens at this point~\citep{mcloughlin2017continuous} to reduce the frequency dimension. The pooling process is discussed further in section~\ref{subsec:pooling}.

Numerous alternative methods of forming spectrograms exist, with the main variants shown in Table~\ref{tab:taxonomy} along with their dimension, element scaling and frequency span. The top three are the linear spectrogram (LS) as described above, followed by variants in which each element of the matrix has been scaled using log, A- or $\mu$-law. The next two variants use Mel and log-Mel scaling, discussed in section~\ref{subsec:melspect}, while the derivations of the bottom three are explored subsequently

\begin{table}[H] 
\caption{Taxonomy of spectrograms.\label{tab:taxonomy}}
\begin{adjustwidth}{-\extralength}{0cm}
\begin{tabularx}{\fulllength}{p{5cm}LLL}
\toprule
\textbf{Description}	& \textbf{Dimensions} & \textbf{Element scale}	& \textbf{Frequency span}\\
\midrule
Linear spectrogram (LS)	& time, frequency (T, F)& scalar $(0,1)$	&  $(0,$ Nyquist$)$\\
Log-scaled spectrogram (LSS)	& T, F & log $(-100$dB$,0)$	& $(0,$ Nyquist$)$\\
A/$\mu$-law	scaling & T, F & log $(0,255)$	& $(0,$ Nyquist$)$\\
\hdashline
Mel-spectrogram (MS)  & T, Mel-F &linear	&  $(0,$ Mel(Nyquist)$)$\\
Log-Mel-spectrogram (LMS)  & T, Mel-F & log $(-100$dB$,0)$	&  $(0,$ Mel(Nyquist)$)$\\
\hdashline
Gammatonegram  (GTG) & trapeziodal/squared-T, F & log	&  $($ERB(0), ERB(Nyquist)$)$\\
Constant-Q transform (CQT)  & trapeziodal/squared-T, F & log/linear	&  $(0,$ Nyquist$)$\\
Stabilised auditory image (SAI)  & non-linear F, lag&  scaled $(0,1)$ &  $(0, 35$ms by default$)$\\
\bottomrule
\end{tabularx}
\end{adjustwidth}
\end{table}

\subsection{Spectrograms are not pictures}
\label{subsec:notpictures}

While spectrograms allow audio and speech to be processed in a deep learning system using techniques that have originally been developed for image processing, caution should be observed for the following three aspects in which spectrograms and picture images differ significantly:

\subsubsection{Colour and greyscale}
Basic linear spectrograms are greyscale with pixel values that are typcially scaled to the range $[0,1]$, but are often colourised for ease of viewing. Colourisation maps  scalar pixel values to RGB values~\citep{CUPbook2016}. In MATLAB$^{\textregistered}$ (The Mathworks Inc.), which is often used for visualisation of spectrograms, each pixel is scaled from 0 to 1. Prior to MATLAB release R2014b, the `Jet' colourmap was used by default to scale a spectrum through blue-green-yellow-orange-red across the range 0 to 1. More recent versions of MATLAB use the `parula' colourmap that scales blue-green-yellow. The popular audio handling tool Audacity maps from -100dB in black, through purple-magenta-light orange to white, for pixel values above -20dB. Both can be modified to display in either greyscale or using other colourmaps.
Python-based tools also impose a colourmap, which may differ based on the library used. In matplotlib, the pcolormap and pcolor functions both default to using the `viridis' colourmap which scales blue-green-yellow across the range 0 to 1. 

While colour scaling produces pretty plots, many researchers simply input a spectrogram, or a spectrogram patch into a convolutional neural network (CNN) that has been designed for image processing, and thus assumes 3 input channels to handle RGB components separately. 
Since the mapping from spectral magnitude to RGB depends arbitrarily on the kind of spectrogram used, and the version of tool used to produce it, there is no logical justification for processing using colour spectrograms. 
Networks such as CNNs can learn a mapping from any scaling, but it may be at the cost of three times the front-end complexity of a greyscale spectrogram input.

\subsubsection{Translation invariance and scaling}
Structures or objects in pictures can very easily be translated to different locations in the image, while remaining the same object. So classification is usually location-invariant. In a spectrogram, translation of structures along the time axis does not change their fundamental nature, but significant translation along the frequency axis can completely change the sound that those structures represent. Unlike in a picture, relationships in the X axis and the Y axis have very different meanings in a spectrogram.

Furthermore, scaling an object in a picture does not change the nature of the object, it only changes its size, i.e. making it appear closer or further away. 
In a spectrogram, scaling a structure that represents a sound event yields a very different result. It adjusts both the time duration of the event and also its frequency span, and has potential to completely change the audible nature of the event.
Importantly, audio deep learning systems need careful matching of scaling between training and inference.

\subsubsection{Local features}

Advanced image processing techniques can exploit both local and global regional characteristics to interpret the content of an image. To do this, systems perform neighbourhood correlations, as well as global texture correlations across an image, and this is part of the motivation behind the use of CNNs~\cite{wang_cnn_2021}. While both local and global correlations are also important in audio tasks such as sound event detection, the nature of those correlations will be very different. For example similar `textures' in frequency ranges of 0-50Hz and 16-18kHz of a spectrogram are unlikely to be significant to understanding the content, whereas in a picture, similar regions of the same texture might be patches of grass at the bottom left and top right of an image -- which also relates to the translation invariance noted above.

Local correlations across the time axis of spectrograms may be more akin to the frame-to-frame difference in video frames than they are to physically proximate points in a picture.

\subsection{Mel-spectrogram}
\label{subsec:melspect}

Mel-spectrograms were inspired by the Mel scale which utilises human equal-loudness data to map frequencies in Hertz to a non-linear scale corresponding to human auditory perception.
The mapping from linear frequency $f_{hz}$  (in Hertz) to Mel frequency  $f_{mel}$~\citep{CUPbook} is generally computed by:
\begin{equation}
\label{eq:mel-spec}
    f_{mel} = 2595.log(1 + {f_{hz}}/{700}),
\end{equation}

A short time Fourier transform (STFT) output vector (i.e. a vector of instantaneous power values for uniformly sampled frequency bins), is transformed into a mel-scale representation vector via a set of bandpass filters, normally with triangular shape.
The bandpass filters are centred at Mel scale frequencies based on eqn.~\ref{eq:mel-spec}. Each triangular filter accumulates the weighted power spectrum sum along the frequency dimension~\citep{CUPbook}. 

Just as a linear spectrogram is constructed from a stack of linear frequency vectors, a Mel-spectrogram is constructed from a stack of Mel-frequency vectors obtained from successive analysis frames.

Since the Mel-spectrogram is based on Mel filters that were developed from human auditory perception experiments, both individual Mel-frequency feature vectors, and their stack into a Mel-spectrogram, has proven effective for various tasks related to human speech analysis.
State-of-the-art systems proposed for Speaker Identification~\cite{speaker_ident}, Speech-to-Text~\cite{whisper-openai}, Speech Emotion Detection~\cite{speech_emotion_detect}, and so on, have used Mel-based spectrograms for the pre-processing feature engineering.

Given that applying Mel filter banks to the SFTF spectrogram across the frequency dimension is effective to capture distinct features in audio signal (i.e., Mel filters are widely used in human speech analysis), several similarly-inspired filter bank representations have also been proposed. These include the Gammatone filter~\cite{gammatone} inspired by cochlea simulation, and the Nearest Neighbour filter~\cite{near_filter} inspired from image pre-processing. Several are illustrated in Fig.~\ref{fig:spectAIM}.

\subsection{Constant-Q spectrogram}
\label{subsec:cqt}
The Constant-Q spectrogram is generated by appplying a constant-Q transform (CQT) which was first introduced in~\cite{cqt_spec} and is closely related to the Fourier Transform.
Like the Fourier Transform, the CQT is formed from a bank of filters, but with the difference that the centre frequencies of each CQT element are spaced in a geometrical tonal space as:
\begin{equation}
\label{eq:cqt-0}
    f_{k} = f_{min}.2^{\frac{k}{b}} ~~~  for  ~~ 1 \leq k \leq K
\end{equation} 
where $f_{k}$ denotes the centre frequency of $k^{th}$, $f_{min}$ is the minimum frequency, $b$ is the number of filters per octave. As the name suggests, the $Q$ value, which is the ratio of central frequency to bandwidth, is constant. It is computed as:
\begin{equation}
\label{eq:cqt-1}
Q = \frac{f_{k}}{\Delta f_{k}} = \frac{f_{k}}{f_{k+1}-f_{k}} = \left (2^{\frac{1}{b}} -1 \right )^{-1}
\end{equation}

In musical analysis, by setting $f_{min}$ and $b$ to directly correspond to musical notes (i.e., choosing $b=12$ and $f_{min}$ as the frequency of, for example midinote 0 or $C_{-1}$), the central frequencies in the CQT will correspond to musical note frequencies, making it effective at capturing musical tones.
As a result, the Constant-Q spectrogram has been widely used for musical analysis~\cite{cqt_ref_01, cqt_ref_02}, but has also been applied to more general sound event detection. Since it is effectively a triangular representation, it is often transformed into a rectangular matrix prior to use.

\subsection{Correlogram}
\label{subsec:corelogram}
The Correlogram utilises autocorrelation to capture the similarity between an audio signal and itself at a given time lag. Autocorrelation vectors are computed for a range of different time lags~\cite{ma2007exploiting}.
Given a long audio signal, it is first separated into short audio segments. 
A correlogram (or auto-correlation vector) is obtained for each audio segment from the auto-correlation coefficients of frequency components along the time axis.
As a result, the long audio signal is presented by a matrix of auto-correlation variables, with each matrix column being an auto-correlation vector representing a given time lag. In structure, it presents similarly to the stabilised audio image (SAI) of Section~\ref{subsec:SAI}.

\subsection{Stabilised auditory image}
\label{subsec:SAI}

Patterson et al.~\cite{patterson1995time} proposed the auditory image model (AIM) in 1995, aiming to simulate the frequency discrimination and amplitude sensitivity of neural activity patterns from hearing. Essentially, an AIM models the function of the basilar membrane, which is part of the organ of hearing within the human cochlea~\citep{CUPbook2016}, when exposed to pure tone.
Walters~\citep{walters2011auditory} integrated this in time (strobed temporal integration, essentally a type of correlogram) to yield the stabilised auditory image (SAI), aiming to improve noise robustness and enhance the detection of periodicity compared to the AIM.

\begin{figure}[tb]
\begin{adjustwidth}{-\extralength}{1cm}
\includegraphics[scale=1]{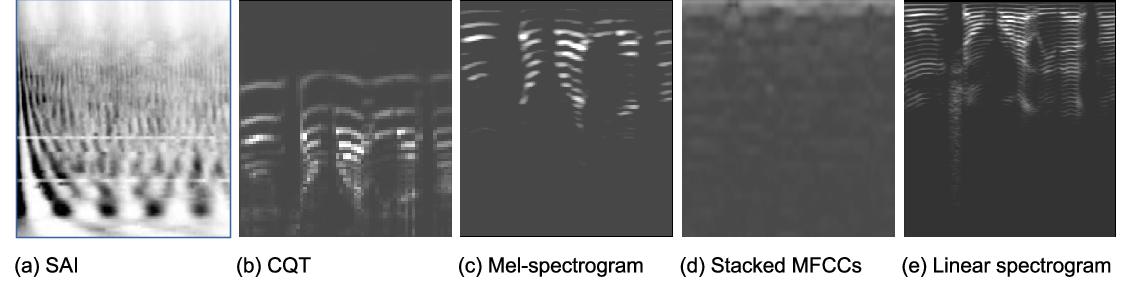}
\end{adjustwidth}
\caption{Illustrations of two dimensional time-frequency spectrograms based on (a) stabilised auditory image, (b) Constant-Q transform, (c) Mel-scaled spectrogram, (d) stacked MFCC, (e) Linear magnitude spectrogram.\label{fig:spectAIM}}
\end{figure}   
\unskip

A single SAI is a two-dimensional representation similar to a classical linear spectrogram, but where the y-axis is frequency and the x-axis represents lag or periodicity. As such it captures the nature of sound in a fixed time window, e.g. 35ms in Fig.~\ref{fig:spectAIM}(a).

SAIs were used as input features in several of Google's early audio recall systems as developed by Lyon et al. \citep{lyon2010sound,lyon2010audio,lyon2011sparse,lyon2011machine}, specifically employing PAMIR (passive-aggressive model for image retrieval), a pre-deep learning ordering algorithm based on statistics obtained from regions within the SAI.
Early attempts at using SAIs with deep learning architectures for sound event detection~\cite{ivmCNNsounddet} were outperformed by linear spectrogram equivalents, probably because of the limited short-time window represented in single SAIs.

\begin{figure}[tbh]
\begin{adjustwidth}{-\extralength}{1cm}
\includegraphics[width=18 cm]{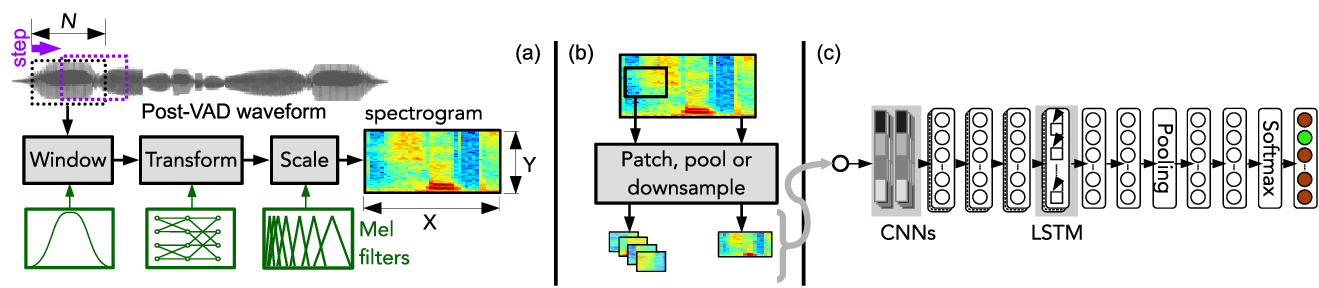}
\end{adjustwidth}
\caption{High level system diagram showing spectrogram features (a) being extracted from an input waveform as a stack of scaled transforms from windowed speech regions then (b) features gathered from patches, pooled regions or a downsampled spectrogram image for (c) input to a deep learning classification pipeline.\label{fig:spec_feature1}}
\end{figure}   
\unskip

\subsection{Patches and regions}
\label{subsec:patches}

Object detection from images, where the characteristic shape of an item can be in any location within the image, as well as any size from small to large, often benefit from techniques where the image is divided into randomly scaled and located patches, each of which are processed independently by the deep learning model~\cite{nowak_sampling_2006}. When the image is a spectrogram, such a process no longer has a physical justification (see 
section~\ref{subsec:notpictures} above). 
The exception is when slices of the spectrogram can be selected in the time domain, spanning the entire frequency range.
However there seems to be limited benefit in allowing the time windows to have different durations, so in practice fixed-sized regions are usually independently inferenced, as in audio spectrum transformer (AST)~\cite{gong_ast_2021}.

\subsection{Scaling and number representation}
\label{subsec:scaling}

Audio samples, such as in the WAVE file format, are typically represented as 16-bit signed linear fixed-point numbers with range $[-32768, 32767]$. During computation they would generally be converted to 32-bit floating point and then divided by $2^{15}$ so they are scaled to a range of $[-1:1)$. An FFT of those samples will, by default, also retain the same 32-bit floating point number format.
Although a true Fourier Transform yields a complex spectrum, spectrograms are usually  formed from the magnitude spectrum, hence each pixel value is always positive, and can be scaled to a range of $[0:1]$.
As mentioned above, log encoding is often used, to provides a perceptually relevant emphasis to the samples. For lower complexity, $\mu$- or A-law is used to convert samples to 8-bit fixed point scaled values in the range $[0,255]$. This can help to substantially reduce downstream computational complexity, at the cost of higher quantization noise.

\subsection{Pooling and downsampling}
\label{subsec:pooling}

State-of-the-art deep learning architectures for performing sound event detection, audio analysis and related tasks including language identification, speaker verification and speech emotion recognition, generally utilise front-end learned layers to compute a one dimensional representation vector (e.g. an embedding), from a raw feature input. 
Thus, whatever input feature is ingested, an intermediate representation -- a fixed dimension embedding -- is produced. A time-stepped series of features yields a time-stepped series of embeddings for analysis.
This stack of embeddings obtained over time can reveal statistics of how the underlying feature, and hence underlying audio signal, varies over time (e.g. over an utterance, or a sound event). The two dimensional block of embeddings from a well-trained front-end is usually amenable for classification.

As we have seen for spectrograms, which are computed from overlapping frames of speech which are windowed, and then transformed to a magnitude spectrum. Magnitude spectra from successive frames are stacked into a two dimensional spectrogram image. This was discussed in Section~\ref{subsec:basic}, and shown in eqns.~\ref{eqn:eqn1} and ~\ref{eq:spectrogramS}.
In early machine hearing systems that pre-dated deep learning approaches, meta-features were extracted from the two-dimensional spectrogram and those features were classified. 
For example, Dennis et al.~\cite{dennis2013image} divided a spectrogram into nine equal-sized regions, and classified the zero and first order statistics from each of the nine regions using SVM. 
Lyon et al.~\cite{lyon2011machine} classified the marginal statistics from rows and columns of an SAI.

The advent of deep learning allowed neural networks to become capable of classifying raw spectrograms directly, but not at full resolution. Thus, downsampling (of samples prior to forming the spectrogram) or pooling of the frequency representation vector (i.e. combining frequency bins) have been common approaches since the very first DNN spectrogram classifier~\cite{ivmDNNsounddet}.

In fact a very similar process happens in many deep learning systems that classify non-spectrogram features too. Examples include MFCC, perceptual linear prediction (PLP) coefficients and filterbank coefficients that are used in tasks such as LID~\cite{Snyder2018,xie2016variance,miao_variance_2021}, discussed further in Section~\ref{subsec:LID}. In almost all cases, one-dimensional features are extracted from overlapping input audio frames and then stacked into a two-dimensional time-frequency block for classification. The frequency-domain features can be pooled at that point, or may have been already downsampled.

\textbf{Pooling} or downsampling along the frequency axis involves taking the mean of, typically, 2, 4 or 8 neighbouring spectral magnitudes to reduce dimensionality by the same factor. In some cases, particularly for MFCC features, in addition to averaging, either max-pooling or standard deviation are computed.

\textbf{Delta-coefficients} are derived in the time-domain to capture changes from one frame to the next. MFCC features are then concatenated with delta-MFCC, and even delta-delta-MFCC features to capture acceleration characteristics~\cite{jinma_odyssey2016}.
Shifted delta cepstral (SDC)  coefficients are commonly used in speech analysis to expand the context window of a classifier. These are formed from a few sequential cepstral delta coefficients per block. For example~\cite{davis1980comparison} concatenated coefficients over 7 blocks, with a shift of 3 between them (called a 7-1-3-7 arrangement). The aim being to capture statistics in a way that mean-pooling in time would not. The same kind of delta computation, shift and concatenation, have also been used with other features like filterbanks and PLPs. The same functionality could be learned within a neural network, particularly a recurrent neural network for time-based changes, but at the cost of additional parameters and training time.

\subsection{Variance normalised features}
\label{subsec:VNF}

Intuitively speaking, when attempting to classify features, the more their statistics differ between two classes, compared to the within-class difference, then the more discriminative the feature is likely to be. This is essentially Fisher's criterion~\citep{malina_extended_1981} restated.
Applying this viewpoint to the downsampling or feature pooling operations used in almost all neural network classifiers (noted above), three of the current authors sought a data-driven approach to maximise Fisher's criterion -- using between-class and within-class variance difference over a development data set -- to identify optimal spectral pooling rules.

Instead of mean pooling fixed blocks of spectral bins (e.g. 8) to reduce the frequency dimension (e.g. from 2048 to 256), the size of the pool is varied across the spectral range based on the variance difference between/within classes. The aim is to normalise the variance contribution of each downsampled feature point. Thus the technique is called variance normalised features (VNF).

Both standard pooling and VNF begin with an identical high resolution spectrum, and aim to reduce the dimensionality before stacking into a spectrogram.
For standard pooling, the low-resolution $N'$ point spectrum $X'(k)$ is obtained from high resolution $N$ point spectrum $X(k)$, where the downsampling factor $D_s=\lfloor N/N' \rfloor $. As noted, it is usually accomplished via mean-pooling;

\begin{equation}
    X'(k) = {1 \over D_s} \sum _{n=kD_s} ^{(k+1)D_s} X(n) \qquad for~ k=0 \dots N'
\end{equation}

Alternatively, max-pooling would be $X'(k) = max\{X_{kD_s} \dots X_{(k+1)D_s}\}$ for $k=0...N'$.

To obtain VNFs, a pre-processing step is required. In that step, the spectrum $X$ is computed over every analysis frame $F$ from all examples in each of $\mathcal{C}$ classes in the development data set.
The bin-wise spectral mean $ \overline{\mathcal{S}_c}$ and variance $\widetilde{\mathcal{S}_c}$ are obtained for each class, $c$, in that set,

\begin{eqnarray} \label{ivm2}
\overline{\mathcal{S}}_c(k) & = & {1 \over F} \sum_{f=0}^{F \in c} X^f(k)\\
\widetilde{\mathcal{S}}_c(k) & = & {1 \over {F-1}} \sum_{f=0}^{F \in c} (X^f(k) - \overline{\mathcal{S}}_c(k))^2
\end{eqnarray}

for all $N$ spectral bins, $0 \le k < N$, and for every class $c \in \mathcal{C}$.

Given the variance and mean spectral characteristics of each class, the per-bin variance is accumulated over all $\mathcal{C}$ classes. This is referred to as the total variance budget,

\begin{eqnarray} \label{ivm3}
V_{c}= \sum_{k=0}^{N} \left| \widetilde{\mathcal{S}_c}(k)- \overline{\mathcal{S}}_c(k)\right| 
\end{eqnarray}

In standard downsampling, the variance contribution of each downsampled point differs depending upon the variance difference across the underlying region. VNF attempts to normalise it so that each downsampled point contributes approximately equal variance difference. This is done by changing from fixed-size pooling regions, with equally spaced partitions, to different sized pooling regions defined by data-driven partition rules.

Those partition rules are computed iteratively from the development set data to specify pooling regions with near-equal amounts of variance contribution. The sum of the variance contributions equals the total budget. One possible partition-setting heuristic is outlined in ~\citep{miao_variance_2021}. Once all partitions are defined, the pre-processing stage has completed.

During operation (i.e. model training or inference) using the VNF pooled features, pooling is applied by obtaining the mean spectral magnitude within each of the pooling partitions. 
The difference between VNF and standard features is that the former uses data-driven pooling regions computed as discussed, whereas the latter use a fixed pooling size to compute all downsized elements.

\begin{table}[H] 
    \caption{Performance of variance normalised features (VNF) compared to standard pooling for three tasks, aiming for higher accuracy and lower C$_{avg}$.}\label{tab:VNF}
    \begin{adjustwidth}{-\extralength}{0cm}
    \begin{tabularx}{\fulllength}{p{1cm}LLL}
    \toprule
    \textbf{Task}    & \textbf{Details}       & \textbf{Fixed pooling}	    & \textbf{VNFs} \\
    \midrule
    SED & 50 class RWCP, 20dB SNR & 94.8\% accuracy & 96.3\% accuracy\\ 
    SED & 50 class RWCP, 0dB SNR & 75.1\% accuracy & 84.0\% accuracy\\
    \midrule
    LID & NIST LRE07 DNN x-vector 3s & 10.17 $C_{avg}$ & 8.80 $C_{avg}$\\
    LID & NIST LRE07 CLSTM 3s & 7.15 $C_{avg}$ & 6.70 $C_{avg}$\\
    \midrule
    DID & Arabic dialect challenge & 3.20 $C_{avg}$ & 2.62 $C_{avg}$\\
    \bottomrule
    \end{tabularx}
   \end{adjustwidth}
\end{table}

The performance of VNFs for three different tasks is shown in Table~\ref{tab:VNF}. Sound event detection (SED) on real-world computing partnership (RWCP) test data~\cite{xie2016variance}, language identification (LID) on NIST Language Recognition Evaluation 2007 challenge data, and dialect identification (DID) for spoken Arabic~\cite{miao_variance_2021}, are performed by models trained from standard fixed pooling inputs, as well as identical models trained with VNF pooled inputs. The aim is for higher accuracy score or lower $C_{avg}$ score. In most tested systems, VNF based pooling outperformed the standard method of mean or max pooling, but it cannot compensate for architectural deficiencies, i.e. it is more important to use a good classifier architecture than to optimise the features, but having found a good classifier architecture, VNF has potential to improve results compared to fixed pooling.

Essentially, any system where spectral bins are mean or max-pooled before classification could potentially benefit from the data-driven VNF approach, as long as a representative development dataset exists from which the one-time pre-processing step can infer suitable partition rules.

\section{Audio analysis}
\label{subsec:sed}

Audio analysis refers to the detection and classification of sounds that lie within the range of human hearing (approx. 20Hz to 20kHz)~\cite{CUPbook2016}. It is related to the field of machine hearing~\cite{lyon2011machine}, which involves endowing computers with the ability to detect and interpret sound in ways analogous to humans.
Generally, we use the term `audio analysis' to refer to non-speech sounds, since speech analysis involves additional techniques which will be considered separately in Section~\ref{sec:speech} -- although there is considerable overlap in the types of features used.

This section will first present a taxonomy of audio analysis, before briefly describing three application areas of audio event detection, anomalous sound detection and the related area of bioacoustics.

\subsection{Taxonomy of audio analysis}

Audio-based classification systems tend to follow a sequential taxonomy as shown in Table~\ref{tab:taxoSED}, although much depends upon the task being performed.
For example in clip detection, or acoustic scene analysis (ASA), a short recording may be analysed, whereas animal call detection in bioacoustics, where segmenting the input into separate animal calls may be difficult, could involve analysing an entire recording.

\subsubsection{Feature extraction} 
Input audio needs to be processed and transformed into features suitable for classification. These could be raw waveform segments~\cite{oord_wavenet_2016}, spectra (including spectrograms), statistical or timbral~\cite{rafsanjani2024unsupervised} features. As we have seen in Section~\ref{sec:taxonomy}, there are many variants of stacked spectra, which could be clipped, segmented, or pooled. Features such as MFCC or perceptual linear prediction (PLP) coefficients can also be stacked to form time-frequency features, as shown in Table~\ref{tab:taxoSED} as `named features'. There may be a natural affinity of certain classification models to particular feature types, but a common alternative is to train a data-driven feature extractor. This is a front-end feature extraction network such as a few CNN layers, that produce features suitable for classification by a back-end classifier~\cite{miao2019new}. The front-end and back-end networks can be trained separately, or end-to-end, if appropriate loss functions can be defined.

Authors also increasingly make use of feature extractors that have been effectively pre-trained by other authors to extract discriminative features for related tasks. Prominent examples include the AST~\cite{gong_ast_2021} as mentioned in Section~\ref{subsec:patches}, PaSST~\cite{koutini2021passt} or HTS-AT~\cite{chen2022hts}. These can be fine-tuned to be used in different tasks, or coupled with domain adaptator layers/blocks~\cite{zheng21_interspeech,PengfeiACMMM2025}.

\begin{table}[tbh] 
\caption{Taxonomy of audio analysis. The input length (first column) can vary from a short frame to a continuous signal, one or more features extracted from this, and then am output class, or timing, obtained.\label{tab:taxoSED}}
\begin{adjustwidth}{-\extralength}{0cm}
\begin{tabularx}{\fulllength}{p{2.2cm}|p{0.1cm}|p{4.4cm}|m{0.4cm}|L}
\toprule
\textbf{Input}	& {} & \textbf{Feature extraction} & \multirow{6}{0.4cm}{\rotatebox{90}{\tt{\small{Stack \& Classify}}}}	&  \textbf{Output}\\
\cline{1-1}
\cline{3-3}
\cline{5-5}
continuous,  & {} & raw waveform/spectrum,         & & one-hot class per instance,\\
full recording, & {} & named features,                 & & posterior probabilities,\\
utterance,   & {} & trained feature extractor or     & & vote over multiple instances,\\
segment or   & {} & pre-trained feature extractor & & average/threshold over time or\\
frame        & {} &                               & & localisation in time\\
\bottomrule
\end{tabularx}
\end{adjustwidth}
\end{table}

The stacked features are then classified by a back-end classifier which outputs typically a one-hot class prediction per instance (e.g. in a detector system), or posterior probabilities in a multi-sound classifier. 
As shown in Table~\ref{tab:taxoSED}, many other possibilities exist for output processing. For example, where a clip of audio to be classified has been split into multiple classification instances, then majority voting, or some kind of weighted averaging provides a single per-class score over multiple classifications for the whole clip. In continuous audio, thresholding of posterior probabilities over a sliding window can yield an activation signal (e.g. for a wake word system~\cite{Chen2024}).
Some tasks are not concerned with clip-level classifications, but require precise detection of the timing of events, in terms of start and end timestamps~\cite{PengfeiACMMM2025}.

A very wide variety of tasks use this basic method of audio analysis from time-frequency spectrogram features. These include sound scene detection (SSD) and auditory scene analysis (ASA)~\cite{bregman1994auditory}, clip recall and recognition~\cite{lyon2010sound}, sound event detection (SED)~\cite{mesaros2021}, anomalous sound detection (ASD)~\cite{zeng2023}, and acoustic classification of speech for purposes such as language identification (LID)~\cite{jinma_odyssey2016}, dialect identification (DID)~\cite{miao2019lstmtdnnconvolutionalfrontenddialect}, speaker identification (SID)~\cite{jiang2019effective}, diarization~\cite{xu2016improved,sun2018speaker}, speaker verification (SV)~\cite{gao2019improving}. There are also medical uses for spectrogram-based auditory analysis that include lung auscultation (stethoscope signal) analysis~\cite{lam2021,9729496}, disease diagnosis from speech~\cite{10.3389/fdgth.2022.886615}, breathing and non-speech vocalisation. This includes from humans~\cite{10.1049/sil2.12233} as well as animals~\cite{kim2025_mdpi}. Music classification or retrieval~\cite{moysis2023}, analysis~\cite{chen2024_mdpi,buisson2024} and even beat tracking~\cite{thapa2024_mdpi} utilise the same basic steps. While spectrogram-based methods either predominate or show excellent performance in most of these fields, alternative approaches exist. Most prominent are those based on direct time-domain waveform analysis~\cite{verma2021audio} as well bag-of-features approaches using statistical indicators~\cite{grzeszick2017bag,lyon2010sound,buisson2024}.

\subsubsection{Overlap and occluded sounds}
In simple environments, targeted sound events may occur in isolation with at most one such occurrence at any time point. This is the basis of clip-level recognition systems which assume that an audio clip contains at most one kind of sound, such as the song of a single bird species, the sound of a gunshot, or a clip from a music recording. However, in complex real-world environments, sound events often coincide with other sounds. 
These include other target sounds (i.e. known sound classes to be recognised) or non-target sounds (i.e. out-of-set sound classes and background acoustic noise). 
The former situation is sometimes referred to as `polyphonic', meaning `many sounds', however the term polyphonic is already used in audio literature to refer to the existence of multiple audio channels -- something that we are not considering here. Almost all sound analysis tasks assume a single channel of audio that potentially captures many sounds. It is thus best described as having ``overlapping or occluded sounds''~\cite{dennis2013overlapping,adavanne2019sound} to avoid confusion with the terminology of multi-channel audio\footnote{If multiple audio channels are available for analysis, this can improve noise removal~\cite{CUPbook2016} as well as enhance localisation and classification performance~\cite{xia2019soundeventdetectionmultichannel}.}. Real-world sounds never occur in isolation, and always have at least some acoustic background noise, so in a sense there are always `many sounds' present in audio collected in the wild.
Research has shown that performance of audio classification systems in even very low levels of noise (i.e. real-world scenarios in a quiet environment), can be very different from the performance with sounds recorded under anechoic conditions of almost zero background noise~\cite{ivmDNNsounddet,ivmCNNsounddet}. Hence real-world deployments of sound classifiers require careful attention to several different techniques that may not always be found in challenge competitions~\cite{alcazar2020classification}.

The temporal occlusion and overlap could be partial or in full. Co-occurring sound events have their frequency-temporal content mixed together. i.e. unlike image occlusion, which implies masking of one object by another, co-incidental audio events are recorded as the linear complex sum of the two events.
A visual inspection of sound mixture spectrograms can sometimes reveal indications of coinciding sounds that are unlike each other (e.g. a long slow low-frequency background during which several short high frequency squeaks or wideband snapping sounds appear). However sounds with similar characteristics can be difficult to discern visually as separate instances, and hence machine learning classifiers or detectors likewise have difficulty in being able to detect similar coinciding sounds~\cite{alcazar2020classification,dennis2013overlapping}.

In general, classification/detection of occluded or overlapping sound events is more challenging than that of isolated ones. In the research literature, three alternative approaches can be taken:
\begin{itemize}
\item \textbf{Recognise from mixed sounds} by implicitly learning from non-occluded sounds as well as all kinds of overlapping sounds. It is also necessary to reframe the classification problem from multi-class to multi-label or 1-vs-rest~\cite{NEURIPS2020_28538c39,mcloughlin2017continuous,BirdNet2021}.
\item \textbf{Separate first} using a source separation framework and then operating classification/detection on the separate channels~\cite{nath2024separation,Sudo2019,Sudo2020}.
\item \textbf{Introduce mixed classes} where potential mixtures are essentially tagged as new combined classes~\cite{Baelde2017,PengfeiACMMM2025}.
\end{itemize}

Separation-based methods are a natural choice for multi-channel audio recordings, in which they can leverage spatial localisation to separate sources, but have shown limited success for single-channel audio. Introducing mixed classes can be useful for commonly mixed sounds, but where there are different degrees and sequences of overlap (e.g. sound 1 occurs first and sound 2 occurs during sound 1, or sound 1 occurs after sound 2, plus more variations), this can negate any benefits of introducing new combined classes. Hence most current approaches are trained in the presence of random mixtures in order to improve robustness, and generality. Interestingly, this may be similar to the robustness benefits gained from the widely-used mixup technique for training classifiers~\cite{10.1007/978-3-030-00764-5_2}.


\subsubsection{Early sound event classification} 
Sound events have a time duration that can range from around a hundred milliseconds (e.g. transient events like \emph{door knocking} or \emph{hand clapping}) to dozens of seconds (like \emph{car passing by} or \emph{baby crying}) or even continual (e.g. \emph{power line hum}). 
Sound classification systems, initially driven by datasets of individually labelled sound clips, were first trained to classify at clip level~\cite{lyon2010sound}. Yet many deployment scenarios involve monitoring of live feeds. Hence the sound event detection task (the identification of \textit{what} sound is present as well as \textit{when} the sound is present -- see Section~\ref{subsec:sed} below), acknowledges that reality. The nature of the task also means that the output from any SED system must come after the sound has ended.

However it is easy to see that sometimes classification needs to occur before a sound event has ended. This is obviously the case for continuous sounds, but also for something like an alarm, or turning off music when someone starts to speak. Some sounds can be classified at frame level, but where the characteristic frequency patterns used to identify a sound have a time span extending beyond a frame (e.g. beyond 10 to 30ms), a different paradigm is necessary.

This was the motivation behind investigations into the timeliness perspective of sound event detection and classification systems. Specifically, the question `how early can a system reliably detect ongoing sound events' from a partial observation of the initial section of an event.
So-called ``early detection'' systems~\cite{phan2015early, phan2017enabling, phan2015early, mcloughlin2018early, zhao2023early} require a classifier to fulfil a monotonicity property on continual input. Early sound event detection systems that scan a sliding window of spectrogram features to dynamically classify segments of input~\cite{zhao2022seed} are particularly useful in surveillance and safety-related applications which require a low latency response from a continuous feed.

\subsection{Sound event detection}
\label{subsec:sed}

Sound Event Detection (SED) aims to identify and temporally localize sound events in audio recordings. It outputs either onset-offset pairs or frame-level activity probabilities for each class~\cite{mesaros2016tut}, and is essential for applications such as environmental monitoring, surveillance and multimedia analysis~\cite{PengfeiACMMM2025}.

Early SED systems used handcrafted features (e.g., MFCC), while recent approaches rely on deep learning with spectrogram inputs. 
This is because spectrograms, especially log-mel representations, are widely adopted due to their alignment with human auditory perception and their ability to capture the time-frequency evolution of sound events~\cite{ivmDNNsounddet,stowell2015detection}. 
CNNs and CRNNs~\cite{cakir2017convolutional,miao2019new} model local correlations effectively, whereas transformer-based models such as PaSST~\cite{koutini2021passt} and HTS-AT~\cite{chen2022hts} better capture long-range dependencies.

Different spectrogram variants have been explored for task-specific benefits. Mel-spectrograms are compact and perceptually motivated; CQT spectrograms suit tonal event detection~\cite{bittner2017deep}; gammatonegrams offer robustness in low SNR~\cite{elias2021gammatone,Pham2021}; and PCEN~\cite{wang2017trainable} improves invariance to background noise through dynamic compression. Overlapping sound events remain a major challenge. As we have noted in Section~\ref{subsec:notpictures}, spectrograms are not translation-invariant, and co-occurring sounds may occupy similar frequency regions. Thus, multi-label classification strategies, attention mechanisms, and source separation methods are often employed~\cite{sed_sep}.

In summary, spectrograms are central to recent SED research due to their compatibility with modern deep models and their descriptive time-frequency structure, though challenges like overlap resolution and domain generalisation persist.

\begin{table}[H]
\caption{Prominent sound event detection methods that utilise spectrogram features, including linear (LS), log-mel (LMS), mel scale (MS), constant-Q (CQT) and gammatonegram (GTG). Tasks include the Real World Computing Partnership (RWCP) sounds, the TUT sound events database, and Domestic Environment Sound Event Detection Dataset (DESED).\label{sed_spect}}
	\begin{adjustwidth}{-\extralength}{0cm}
		\begin{tabularx}{\fulllength}{p{0.9cm}p{0.7cm}p{4.9cm}LLL}
\toprule
\textbf{Year} & \textbf{Ref.}	& \textbf{Task}	& \textbf{Spectrogram type}     & \textbf{Resolution \& settings} & \textbf{Pooling}\\
\midrule 
2014 & \citep{ivmDNNsounddet} & 50 class RWCP & LS & $[30{\times}24]$, 16kHz & vote\\
2017 & \citep{Adavanne2017} & 6 class TUT events~\cite{Heittola2017}$^1$ & LMS+LS & $[240{\times}256]$, 44.1kHz & max\\
2019 & \citep{Lin2019} & 10 class from~\cite{Audioset}+~\cite{Fonseca2017freesound} & LMS & $[64{\times}500]$, 16kHz  & median\\
2020 & \citep{ivm_timefrequency_2020} & 50 class RWCP & LS,GTG,CQT & $[52{\times}40]$, 16kHz  & mean\\
2022 & \citep{Ebbers2022} & 10 class DESED~\cite{turpault:hal-02160855}$^2$ & LMS & $[128{\times}960]$, 16kHz  & mean\\
2023 & \citep{Kim2023} & 10 class DESED~\cite{turpault:hal-02160855}$^2$ & MS & $[128{\times} 1001]$, 16kHz  & mean\\
2024 & \citep{Schmid2024} & 11 class DCASE24 task 4~\cite{Martinmorato2023} & MS & $[128{\times}100]^3$ in AST+fPaSST, 16kHz  & ensemble-mean\\
\bottomrule
\multicolumn{6}{l}{$^1$ DCASE 2017 task 3}\\
\multicolumn{6}{l}{$^2$ DCASE 2022/3/4 task 4}\\
\multicolumn{6}{l}{$^3$ In 16x16 patches}\\
\end{tabularx}
	\end{adjustwidth}
\end{table}

Table~\ref{sed_spect} samples approximately a decade of advances in the SED field since the first published use of spectrograms with deep learning~\cite{ivmDNNsounddet}. Numerous spectrogram variants have been applied to this field, with many recent systems favouring Mel spectrograms or log-Mel spectrograms. The common evaluation tasks have largely been driven by DCASE (Detection and Classification of Acoustic Scenes and Events) challenges and workshops\footnote{DCASE (Detection and Classification of Acoustic Scenes and Events): \url{https://dcase.community/}}, which has also resulted in a tendency towards datasets with a relatively small number of classes. Finally we note that embeddings from pre-trained transformers operating on spectrogram inputs, as mentioned above, can be utilised to improve performance, while effectively reducing the overall training resource required through model re-use and adaptation.

\subsection{Anomalous Sound Detection}

Anomalous Sound Detection (ASD) is the task of determining whether a given audio signal contains abnormal or anomalous sounds that deviate from patterns typically observed under normal conditions~\cite{kim2025_mdpi,TANG2023113294,zeng2023}. This task is of significant importance in scenarios such as industrial monitoring~\ and security surveillance. For instance, in factory environments, the early detection of machine failures, system anomalies, or unexpected environmental events can effectively prevent accidents, reduce downtime, and enhance overall safety~\cite{qurthobi2022}.

Unlike conventional audio classification tasks, ASD is typically conducted in an unsupervised manner, as anomalous events are rare and diverse, making them difficult to define and annotate in advance during training. As a result, most methods rely on modeling the distribution of normal sounds, and identify anomalies by evaluating reconstruction errors, likelihood scores, or deviations in feature embedding space~\cite{zeng23b}.
For example, in DCASE 2025 Challenge Task 2 (and similar to 2020-2024 task 2), only normal audio data is provided for training. Models are required to learn the feature distribution of normal audio samples and perform classification during testing by comparing the characteristics of normal and abnormal audio samples.

Current ASD systems can be broadly categorised into two main approaches: generative, and discriminative. Generative methods, grounded in the paradigm of self-supervised learning, tend to employ autoencoder-based models, like AE~\cite{AE_ASD}, VAE~\cite{9054344}, PAE~\cite{PAE_ASD}, to learn the feature distribution of normal audio. Anomalies are identified by computing a reconstruction error between the generated and original samples. The underlying assumption is that normal samples result in low reconstruction errors, whereas anomalies yield significantly higher errors. However the strong generalisation ability of generative models, even in mismatched domains, means that they can be capable of reconstructing some anomalous samples, leading to false negatives~\cite{zeng23b}.

Recent discriminative learning methods use an Outlier Exposure (OE) strategy~\cite{OE_ASD}. In this case, additional meta-information obtained during the data collection process (e.g. machine ID and attributes, such as operating condition) is utilised to train a classifier on the acoustic features of normal samples. Normal samples from different categories are treated as pseudo-anomalies relative to the target category. A compact normal feature space is then constructed using both this meta-information and a feature extractor, such as ResNet~\cite{Resnet_ASD}, MobileFaceNet~\cite{STgram}, or Transformer-based models~\cite{large_scale_model_in_ASD, anopatch}. During the inference phase, the feature distance between a test sample and the normal training samples is regarded as a proxy indicator of the degree of anomaly.

\begin{table}[H] 
    \caption{Spectrograms used in ASD, primarily log-Mel (LMS, their resolution, sample rate and scaling.). \label{tab:task_asd_spectrograms}}
    \begin{adjustwidth}{-\extralength}{0cm}
    \begin{tabularx}{\fulllength}{p{4cm}LLLL}
    \toprule
    \textbf{Method}    & \textbf{Spectrogram}       & \textbf{Pixels}	    & \textbf{Sample Rate} & \textbf{Scale} \\
    \midrule
    Chakrabarty et al.~\cite{7471668}     & LMS & 	128 $\times$ T     & 8kHz & log \\
    Zeng et al.~\cite{zeng23b} & LMS & 	128 $\times$ T    & 16kHz & inverted log \\
    Li et al.~\cite{li2024machine}	& LNS & 128 $\times$ T	& 16kHz & log \\
    Liu et al.~\cite{STgram} & LMS+Tgram & 128 $\times$ T	& 16kHz & log \\
    Yin et al.~\cite{10890695} & LMS & 128 $\times$ T	& 16kHz & normalised \\
    \bottomrule
    \end{tabularx}
    \end{adjustwidth}
\end{table}

Table~\ref{tab:task_asd_spectrograms} summarises representative spectrogram-based approaches used in ASD, where LNS denotes log non-uniform spectrum. Chakrabarty et al.~\cite{7471668} were the first to apply spectrograms for anomalous detection, utilizing log-mel spectrograms  (LMS) with 10-frame concatenation as input to Restricted Boltzmann Machines. 
Zeng et al.~\cite{zeng23b} employed log-mel spectrograms with transposed filters, where filters are sparse in the low-frequency region and dense in the high-frequency region. For machine sounds, high-frequency components often contain richer and more discriminative information, while the low-frequency part tends to be more noise-prone.
Li et al.~\cite{li2024machine} further advanced this direction by computing F-ratios to analyze the distribution of information across the spectrum and designed machine-specific non-uniform filterbanks. Recently, more studies in ASD have begun combining both spectrogram and time-domain information as input features. 
For example, Liu et al.~\cite{STgram} proposed the STgram-MFN method, which concatenates temporal features with a log-Mel spectrogram for classification. The time-domain features (called a `Tgram') are derived from a trained CNN network, with an ArcFace-derived loss. Interestingly, this revealed that the Tgram feature was able to provide useful, and complementary, information alongside the log-Mel spectrogram.
Taking a different approach, Yin et al.~\cite{10890695} applied a diffusion model to synthesize log-mel spectrograms for data augmentation, achieving a state-of-the-art (SOTA) result with an official score of 67.12\% in DCASE 2024 Challenge Task 2 using a discriminative approach.

Despite the promising performance of spectrogram-based methods in ASD, several challenges remain. First, spectrograms are sensitive to noise and machine type, requiring tuning. This includes different parameter settings for different machine categories, which limits model generalisation. Second, under domain shift scenarios, reconstruction-based spectrogram methods may fail to detect anomalies due to misleadingly low reconstruction errors. Finally, in the absence of machine metadata, discriminative models struggle to construct a compact normal sound space, resulting in significant performance degradation. There is thus significant potential for ongoing research in this area.

\subsection{Bioacoustics}
\label{sec:bio}

Bioacoustics is primarily applied in three core tasks: \textit{species classification}, \textit{call segmentation}, and \textit{sound event detection}. The purpose of these tasks is to automate the analysis of animal vocalizations recorded in natural environments. \textit{Species classification} involves identifying a species from audio recordings.~\cite{stowell2022computational, tosatodeep, heinrich2024audioprotonet}. \textit{Call segmentation} aims to isolate individual vocalisations (e.g., bird syllables, frog calls, whale units) within recorded audio streams \cite{stowell2022computational, hershey2015deep}.

We also note increasing research relating to animal vocalisations for purposes such as health monitoring~\cite{kim2025_mdpi}, emotion recognition~\cite{dang2025} and potentially communications. This includes human-to-animal voice conversion techniques such as ``Speak like a dog''~\cite{suzuki2022speak} or wider species-to-species conversion using feature fusion that includes spectrograms~\cite{kang2025humans}.

Animal call segmentation, by identifying \textit{when} and \textit{which} biological sounds occur is a subset of Sound event detection (SED). This is characterised by long, noisy recordings, often with multiple overlapping species~\cite{stowell2022computational}. Each of these tasks presents distinct challenges. For classification, models must differentiate highly similar calls across species, sometimes with very few labeled examples~\cite{tosatodeep}. Segmentation is complicated by overlapping sounds, variable call durations, and background noise~\cite{stowell2022computational, hershey2015deep} (sometimes with the background noise inextricably correlated to the species). Detectors and classifiers must deal with complex real-world soundscapes, requiring highly robust models to generalise across time and different environments. Despite these difficulties, recent spectrogram-based approaches have shown strong performance across all three main task categories~\cite{stowell2022computational, heinrich2024audioprotonet}. 

\begin{table}[H]
\caption{Deep learning front-ends in bioacoustic analysis that make use of raw waveforms (top three), linear and log-Mel spectrograms (LS, LMS) and stabilized auditory image (SAI) (middle four) and hybrid approaches (bottom two).\label{tab:frontends}}
\begin{adjustwidth}{-\extralength}{0cm}
\begin{tabularx}{\fulllength}{>{\raggedright\arraybackslash}X 
                             >{\raggedright\arraybackslash}X 
                             >{\raggedright\arraybackslash}X 
                             >{\raggedright\arraybackslash}X}
\toprule
\textbf{Technique} & \textbf{Input Type} & \textbf{Task(s)} & \textbf{Taxa} \\
\midrule
SincNet \cite{ravanelli2018sincnet} & Raw waveform & Species classification & Birds \\
SampleCNN \cite{lee2017samplecnn} & Raw waveform & Music auto-tagging & Music \\
RawNet \cite{jung2019rawnet} & Raw waveform & Speaker verification & Humans \\
\hdashline
CNN/ResNet \cite{stowell2022computational, bravosanchez2021bioacoustic} & LMS & Species classification, SED & Birds, Frogs, Whales \\
PCEN-enhanced CNN \cite{wang2017trainable, zeghidour2021leaf, allen2021cnn} & PCEN-Mel & Low-SNR event detection & Birds, Whales \\
CNN on STFT \cite{hexeberg2023semi} & LS & Call segmentation & Bats \\
CNN with spectrogram and stabilized auditory image input \cite{ivmDNNsounddet} & LS + SAI & Sound event classification & General sounds) \\
\hdashline
LEAF \cite{zeghidour2021leaf} & Learned spectrogram from raw waveform & Species classification, detection in noise & Birds, Whales \\
Wavegram-Logmel-CNN \cite{kong2020panns} & Wavegram + LMS & General classification & Various \\
\bottomrule
\end{tabularx}
\end{adjustwidth}
\end{table}

Table \ref{tab:frontends} summarizes recent trends in bioacoustic deep learning, highlighting a strong preference for spectrograms that are either linear (LS) or log-Mel spectrograms (LMS). The PCEN (per-channel energy normalisation) enhanced Mel spectrograms address background noise by essentially performing scaling and auto-gain control on each frequency bin~\cite{PCEN2019}. However the baseline linear spectrogram remains relevant in segmentation tasks—particularly in high-frequency domains such as bat echolocation—due to its simplicity and fine temporal resolution~\cite{hexeberg2023semi}. It is also true that the Mel scale, based on human hearing, would be inappropriate for spectrograms encompassing the ultrasonic region.
At audible frequencies, log-Mel spectrograms remain widely adopted for species classification, given their well-demonstrated predictive performance when paired with CNNs. For example, on a bird classification task, a ResNet50 trained on Mel spectrograms achieved an accuracy of 0.77 and ROC AUC of 0.80, outperforming a raw waveform CNN baseline, which yielded 0.71 and 0.76, respectively~\cite{bravosanchez2021bioacoustic}. 
PCEN-enhanced spectrograms have gained traction for their robustness in noisy settings, enabling better detection of faint or distant calls across taxa~\cite{zeghidour2021leaf, stowell2022computational}. Although hybrid models such as LEAF attempt to combine the strengths of raw waveforms and spectral representations, real-world applications still show a clear advantage for spectrogram-based features, which have been shown to be robust, interpretable, and scalable across taxa and datasets.

Spectrogram-based methods are not without limitations, however. In species classification, they often struggle with fine-grained distinctions between species that produce acoustic calls in overlapping frequency bands~\cite{hexeberg2023semi, stowell2022computational}. Variability in vocal structure, such as regional `dialects' or age-related changes, can also reduce accuracy. One structural limitation may be the use of fixed-size spectrogram windows that can truncate short calls, or blend closely spaced vocalisations in recordings with many overlapping signals~\cite{stowell2022computational, hershey2015deep}. 
This is also problematic for insect chirps and brief bat calls. 
Although preprocessing using PCEN improves robustness to loudness variation and background noise~\cite{wang2017trainable, zeghidour2021leaf}, it does not fully resolve these segmentation and overlap challenges. Across all tasks, significant training challenges exist due to class imbalance, sparse labels for rare species, and taxonomic bias in training data. These issues limit generalisation and deployment, especially when extending models to new ecosystems or to poorly studied taxa~\cite{stowell2022computational} for which, ironically, they may be most needed.

\section{Speech analysis}
\label{sec:speech}

Speech analysis differs from pure auditory analysis due to the linguistic and semantic nature of the underlying speech signal. It not only conveys different information, but our understanding of it (i.e. labels) has greater complexity and allows more resources to be applied for speech analysis, compared to general audio analysis.
Speech analysed at frame-level primarily captures acoustic properties of the human vocal system during production of the current senone or phonetic unit~\cite{jinma_odyssey2016}, or non-verbal vocalisation. It provides a snapshot view of both how the speech is being produced, which reveals information about the speaker, as well as the nature of what is being produced, which reveals information about the current pronunciation unit or sound.

Speech analysed at utterance level captures linguistic content, which reveals information about the semantic meaning conveyed by the speech. Dynamics of frame-level changes also reveal information about the speaker, including their identity, their mood, gender, age, as well as potentially reflecting several physical and mental conditions. The analysis often makes use of the statistical variation in time as the frame-level features evolve~\cite{jinma_odyssey2016}. Stacking frequency-domain frame-level features creates a time-frequency image, which is a type of spectrogram.

Considering the audio analysis framework taxonomy of Fig.~\ref{tab:taxoSED}, front-end features tend to be at frame level, whereas the back end classification tends to be utterance level. Many variants to this simple understanding exist, such as front-end features spanning several frames, or word-level, chunk-level~\cite{song2025efficienttransferlearningmethod} and entire recording analysis.

The following subsections survey three specific speech tasks that encompass that range, albeit with different objectives: language and dialect identification, speaker verification and speech emotion recognition.

\subsection{Language and dialect identification}
\label{subsec:LID}
The objective in language and dialect identification (LID/DID) is to extract information from recorded speech utterances -- often with different linguistic content, spoken by different speakers, and captured in varying acoustic environments -- and to develop methods that can reliably determine which language or dialect is spoken, typically from among a closed-set of known alternatives. While many approaches have been proposed over the years, state-of-the-art systems rely on the evidence that acoustic features carry robust language-specific cues that are suitable for front-end feature extraction~\cite{OSHAUGHNESSY2025}. Extracted embeddings are then typically stacked and classified as noted above, using deep neural network-based back-end architectures.

The well established MFCC~\cite{davis1980comparison} features, representing the short-time power cepstrum of speech, mapped onto a Mel scale -- essentially the discrete cosine transform (DCT) of log-Mel filterbank features. They involve weighted pooling across overlapping spectral regions (i.e. Mel coefficients).
Because MFCC are extracted from short (e.g. 25 ms) frames, they are limited in their ability to capture longer-term temporal dependencies in speech, hence are typically stacked with their delta and delta-delta (framewise difference, and difference between framewise differences) across an utterance. This captures first- and second-order temporal derivatives to improve the modeling of speech dynamics~\cite{miao2019new,miao2021d,liu2022efficient,dey2024towards}. 
 Shifted delta coefficients~\cite{kohler2002language}, as discussed in Section~\ref{subsec:pooling}, similarly help to capture patterns that extend beyond individual frames, which is important for modeling the sequential nature of speech.

More recently, research has moved beyond handcrafted features and instead demonstrated the effectiveness of directly using raw log-Mel spectrograms as input~\cite{cai2018novel,alumae2023exploring}. The fact that spectrograms better preserve the time-frequency structures of speech, enables convolutional or recurrent neural networks to learn discriminative representations from the data, without relying heavily on engineered features.

\begin{table}[H] 
    \caption{Prominent LID research showing various kinds of spectrogram, including linear (LS) and log-Mel spectrograms (LMS). EER is equal error rate.\label{tab:task_lid}}
    \begin{adjustwidth}{-\extralength}{0cm}
    \begin{tabularx}{\fulllength}{p{3cm}p{4cm}p{5.5cm}L}
    \toprule
\textbf{Method} & \textbf{Spectrogram}  & \textbf{Resolution} & \textbf{Task}        \\
    \midrule
Ma et al.~\cite{jin2018lid} & PLP + bottleneck& 48 $\times$ 21 & 23 languages, EER 4.38\%$^1$\\
Kaiyr et al.~\cite{Kaiyr2021} & LS, CNN-RNN & 116$\times$200 5-10s segments & 7 languages, acc. 94.3\%$^2$\\
Liu et al.~\cite{liu2022efficient} & MFCC+delta+delta-delta & 39+39+39, 25ms window, 10ms hop & 14 languages, EER 3.82\%$^3$\\
Miao et al.~\cite{miao2021d} & MFCC+D-MONA & 23$\times$5 frames & 14 languages, EER 1.15\%$^3$\\
Tjanda et al.~\cite{tjandra2022improved} & LMS & 80$\times$4, 25ms window, 10ms hop & 26 languages, acc. 90.3\%$^4$\\
\bottomrule
\multicolumn{4}{l}{$^1$ Evaluated on 10s utterances using NIST LRE2009.}\\
\multicolumn{4}{l}{$^2$ Evaluated on 5-10s clips.}\\
\multicolumn{4}{l}{$^3$ Evaluated on 10s utterances using NIST LRE2017.}\\
\multicolumn{4}{l}{$^4$ Evaluated on 6s utterances.}\\
    \end{tabularx}
    \end{adjustwidth}
\end{table}

Table~\ref{tab:task_lid} presents some representative works for LID. Evaluation tasks vary widely in terms of the number of languages, the utterance length and the variety of speakers. 
Features also vary between approaches, and clearly there are trade-offs between feature size and resolution.
Performance is generally measured by Equal Error Rate (EER), or accuracy (and by C$_{avg}$ in newer works).
In general, the more languages and the shorter the evaluation clip, the more difficult the task becomes. Confusion matrices~\cite{jin2018lid} reveal that inclusion of similar languages can significantly reduce average performance scores. All of the systems in Table~\ref{tab:task_lid} contain deep recurrent networks, hence even input features with a short context length are able to benefit from time-domain context within the network to perform well.

\subsection{Speaker verification}
Speaker verification (SV) is the task of determining whether an input speech signal matches a claimed speaker identity from a set of enrolled speakers. SV serves as a fundamental component in biometric authentication, forensic analysis, and secure access control systems -- for example, enabling user verification for banking transactions and for unlocking mobile devices~\cite{Reynolds2000,Dehak2011}. The related `speaker validation' task computes the probability that a given speaker is who they claim to be, while `forensic speaker identification' aims to discern as much information as possible, including identity, of an unknown speaker.

Unlike most automatic conventional speech classification tasks, SV operates in an open‑set setting because speakers being analysed may have not been seen during training. However training can be conducted in a supervised manner, since labeled speaker data are readily available for enrolment. During training, models learn discriminative representations of target and non‑target speakers, while during inference they compare embeddings of test utterances against enrolled models to make acceptance or rejection decisions~\cite{Variani2014,Snyder2018}.

In benchmarks such as the NIST Speaker Recognition Evaluations (SRE) and the VoxCeleb challenges, thousands of labeled utterances recorded over telephone (8kHz) and ``in the wild'' (16kHz) conditions are provided for system development. Evaluation is performed by scoring the similarity. This often uses cosine distance or probabilistic linear discriminant analysis (PLDA) between enrolment and test embeddings~\cite{liu2024dp}. Performance is generally measured by Equal Error Rate (EER) and Detection Cost Function (DCF) \cite{Nagrani2017}.

Current SV systems can be broadly categorised into generative embedding methods and discriminative embedding methods. 
Generative embedding methods, such as GMM‑UBM and total variability (i‑vector) frameworks, model speaker and channel variability via statistical supervectors and employ PLDA for scoring \cite{Reynolds2000,Dehak2011}. Discriminative embedding methods leverage deep neural networks to directly learn fixed‑dimensional speaker embeddings: the d‑vector approach averages frame‑level DNN activations \cite{Variani2014}, the x‑vector architecture uses TDNNs with statistics pooling \cite{Snyder2018}, and enhanced variants like ECAPA‑TDNN incorporate channel attention and hierarchical feature aggregation to further improve robustness \cite{Desplanques2020}.

\begin{table}[H] 
    \caption{Prominent SV research showing a progression of spectrogram use, where LMS refers to log-Mel spectrogram and FB are filterbanks. \label{tab:task_sv_spectrograms}}
    \begin{adjustwidth}{-\extralength}{0cm}
    \begin{tabularx}{\fulllength}{LLL}
    \toprule
    \textbf{Method}                       & \textbf{Spectrogram}            & \textbf{Resolution}        \\
    \midrule
    Reynolds et al.~\cite{Reynolds2000}          & MFCC+context                            & 13 $\times$ T         \\
    Dehak et al.~\cite{Dehak2011}                & i-vector from 60d MFCC         & 200 WCCN           \\
    Variani et al.~\cite{Variani2014}            & trained from 40d FB+context        & 256 d-vector     \\
    Snyder et al.~\cite{Snyder2018}              & 60d MFCC+delta+delta-delta           & 150 x-vector     \\
    Desplanques et al.~\cite{Desplanques2020}    & LMS            & 80 $\times$ 80                        \\
    Liu et al.~\cite{liu2024dp}				  & LMS            & 128 $\times$ 304    			\\
    \bottomrule
    \end{tabularx}
    \end{adjustwidth}
\end{table}

Table~\ref{tab:task_sv_spectrograms} summarises several representative works for SV, including recent spectrogram-based feature representations. Despite the promising performance of these methods, several challenges remain. First, channel and domain mismatch between training and test recordings leads to performance degradation under cross-corpus and cross-device conditions~\cite{Dehak2011}. Second, short‑duration utterances often yield unreliable embeddings, increasing error rates. Finally, spectrogram parameter tuning for different languages, noise environments, and recording devices remains a manual and time-consuming process. The need to re-tune systems for a new task, coupled with limitations in both robustness and generalisation hinders the large-scale deployment of this technology~\cite{Campbell2006}. It also presents opportunities for future research.

\subsection{Speech emotion recognition}
Speech Emotion Recognition (SER) seeks to infer affective states from analysis of speech. It generally models prosodic, spectral, and temporal variations correlated with arousal and valence, as well as with discrete emotion categories~\cite{schuller2009}. In contrast to speaker verification, a task that benefits from representations that are stable across a speaker’s vocal space, SER requires \textit{speaker-invariant} embeddings that encode emotion-related acoustic fluctuations. 
Much past research has indicated that emotional expressions often correlate with changes in fundamental frequency, harmonic-to-noise ratio, spectral tilt, and the bandwidth of formants. 
Spectrograms have gained widespread use in SER research in recent years, given that their time-frequency viewpoint into evolving speech signals.
Unlike raw waveform inputs, spectrograms reveal clear energy distributions across frequency bands over time, and these can be used to describe affective correlates.

Early SER systems generally employed hand-crafted descriptors such as MFCC, pitch- and energy-based prosodic features, as well as jitter/shimmer measures. These would often be combined into utterance-level statistics, e.g. openSMILE~\cite{eyben2010opensmile,schuller2011}. While these features have been shown effective for acted emotional speech (i.e. databases of actors representing emotions on demand), they inherently compress some spectral detail. 
For example, MFCC decorrelate and smooth spectral envelopes using a DCT. High-frequency cues, which have been associated with emotional arousal, can be lost or misaligned in the process. 
Utterance-level pooling further removes useful information regarding temporal evolution, which limits the ability to capture brief spectral features, and may hide discriminative distributions. 
This probably contributes to a widely observed performance degradation in which models trained using one corpus perform much less well when they are evaluated using another, i.e. cross-corpus evaluation, or generalisation testing~\cite{parry2019analysis}.

The transition to explicit use of spectrograms enabled richer modelling of time-frequency features. Linear, log-Mel and Mel-spectrograms preserve local spectral–temporal patterns, allowing convolutional neural networks (CNNs) to learn filters sensitive to emotion-associated frequency characteristics as noted above~\cite{satt2017}. 
Time domain cues that unfold over several successive frames can be effectively captured by recurrent architectures like BLSTMs. For example, the evolution of pitch and intensity profiles and spectral modulation over time~\cite{trigeorgis2016adieu}. Attention mechanisms further refine the time-domain sensitivity by weighting frames that contribute more strongly to emotional perception, while down-weighting linguistically dominant or neutral segments~\cite{mirsamadi2017}. Importantly, spectrogram configuration choices such as window length, hop size, and number of Mel bands can affect the emotional cue representation granularity in both time and frequency dimensions. Smaller hop sizes, allied with a recurrent network, increase sensitivity to rapid prosodic change, whereas higher Mel-band resolution allows finer modelling of high-frequency structures. As with other spectrogram-based audio analysis tasks, there are trade-offs to be made between granularity and context, especially in the time domain.

As with SV and, to some extent LID, recent SER systems integrate high-resolution spectrograms with deep sequence models such as CNN–Transformer hybrids. Transformers capture long-range dependencies and global contextual structures. This can complement the ability of CNNs to extract localised patterns. Meanwhile self-supervised learning (SSL) models such as wav2vec 2.0, HuBERT, and WavLM provide contextualised frame-level embeddings that have been learned from large unlabeled speech corpora~\cite{pepino2021,chen2022wavlm}. Although WavLM operates on raw waveforms, there is some evidence that intermediate layers can encode spectrally-relevant information, such as fundamental frequency trajectories, and amplitude modulation patterns -- attributes that overlap with those observable in spectrograms~\cite{diatlova2024}. As a result, SSL embeddings can serve as either an alternative to, or a complement for, spectrogram features in modern SER pipelines. 

\begin{table}[H]

    \caption{Representative SER systems illustrating the evolution from hand-crafted features to high-resolution spectrograms and SSL-based embeddings. FB refers to filterbank, LMS is log-mel spectrogram. \label{tab:task_ser_spectrograms}}

    \centering 
    \begin{adjustwidth}{-\extralength}{0cm}
    \begin{tabularx}{\fulllength}{p{3.6cm} p{5.5cm} L}
        \toprule
        \textbf{Method} & \textbf{Feature Type} & \textbf{Representation} \\
        \midrule

        Schuller et al.~\cite{schuller2011} & MFCC + prosody + energy
            & 1582-d openSMILE \\

        Satt et al.~\cite{satt2017} & LMS
            & $\sim$40--64 Mel bands, 25ms window, 10ms hop \\

        Mirsamadi et al.~\cite{mirsamadi2017} & FB with frame attention
            & 40dim FB, 25ms window, 10ms hop \\

        Trigeorgis et al.~\cite{trigeorgis2016adieu} & LMS + channel attention
            & 40dim FB with 40ms frame, 5ms hop \\

        Pepino et al.~\cite{pepino2021} & wav2vec 2.0 SSL embedding
            & 768dim contextual frames \\

        Chen et al.~\cite{chen2022wavlm} & WavLM SSL embedding
            & 1024dim contextual frames \\

        Chowdhury et al.~\cite{chowdhury2025} & LMS + 5 other features
            & 64dim LMS 20-30ms, and 126dim other features \\

        \bottomrule
    \end{tabularx}
    \end{adjustwidth}
\end{table}

Table~\ref{tab:task_ser_spectrograms} summarises representative SER approaches and highlights the shift from coarse statistical descriptors to high-resolution spectrograms, and to spectrogram-informed latent embeddings from pre-trained models.

Despite many recent advances, several challenges persist in SER research using spectrograms. Firstly, emotional correlates vary across speakers, languages, speaking styles, as well as recording conditions. Spectral tilt, harmonic structure as well as prosodic patterns are inconsistent across corpora, which remain highly influenced by their recording conditions, task (e.g. spontaneous, scripted etc.) and labelling methodology. This contributes to domain mismatch~\cite{amjad2025}. 
Secondly, many emotion-relevant cues occur at short temporal scales that are sensitive to spectrogram settings, where inappropriate frame sizes, windowing or Mel resolution may obscure rapid spectral transitions. Other emotion cues may evolve over a long timescale that is sensitive to the recurrence length or context size of features.
Third, while SSL features appear to offer robustness, their lack of explicit frequency structure makes it difficult to model multi-resolution emotional cues, or to interpret how spectral information influences predictions. Addressing these issues may require frameworks that can integrate interpretable time-frequency structures obtained from spectrograms with the robustness and abstraction provided by waveform-based analysis.

\section{Conclusion}
\label{sec:conc}

This paper has surveyed the nature and application of spectrogram time-frequency features when used for audio and speech analysis. 
Beginning with the definition of a spectrogram, we considered element scalings such as Mel, log-Mel, A-law and $\mu$-law, as well as alternaive transforms including Gammatonegrams, stabilised auditory images (SAI) and constant-Q tranforms. Spectrograms formed from stacking other kinds of frequency-vectors in time, such as MFCC, PLP, filterbanks and embeddings from pre-trained models. Settings including frequency resolution, time span, range, frame size and hop were considered alongside the related task of pooling or downsampling resolution, and the need to vote, or otherwise process indvidual frames to obtain a per-utterance/chunk/recording classification, as well as timestamps for start and end of events, where detection is required.

In each of the analysis domains sampled within this paper (namely SED, ASD, bioacoustics, LID/DID, SV and SER), the past decade has seen a shift away from statistical features (e.g. variance, shimmer, skewness), through handcrafted features such as MFCC and filterbanks, to spectrograms. Early spectrogram-based deep learning classifiers~\cite{ivmDNNsounddet, ivmCNNsounddet} used small rectangular patches. This was because limited training datasets restricted the model complexity that could be effectively trained, which in turn limited the size of input features. As more training data became available, larger models were possible and spectrogram patch sizes tended to increase. Recurrent networks enabled deep neural networks to exploit time-domain statistics, allowing the size of spectrogram patches to reduce -- or to reduce in time span but increase in frequency resolution (often through larger analysis windows with smaller hop sizes).

While spectrogram features have been shown effective in many audio analysis domains, pooling to downsample tends to obscure fine detail. Variance normalised features (VNF) were proposed to define more nuanced pooling rules to maximise between-class variance compared to within-class variance. However the complexity of optimising front-end features, and the need for a large representative dataset has led researchers increasingly to  adopt pre-trained foundation models. These have often been well trained for a different but related task, such as automatic speech recognition, where large high quality datasets are readily available. For speech systems, models are often pre-trained as ASR phone detectors. For audio, models such as AST~\cite{gong_ast_2021}, PaSST~\cite{koutini2021passt} and HTS-AT~\cite{chen2022hts}, they are trained as classifiers using large scale audio datasets.
Adaptation methods or fine tuning are then used to harness these models for target tasks such as SER~\cite{chen2022hts}, SV~\cite{Chen2023_SV} and more using pre-trained speech models, or SED~\cite{PengfeiACMMM2025}, bioacoustics~\cite{ghaffari2025robust} and more for audio models. In each, the performance of the adaptation techniques is crucial to the resulting system performance.
Apart from reasons of training efficiency, a strong motivation for use of pre-trained models is the example of the human auditory system -- a fixed external capture system (pinna, outer and middle ear), a fixed feature transformer (inner ear, auditory nerve)~\cite{CUPbook2016}, both of which handle all auditory tasks. However there are specialised back-end processes operating within physically separate areas of the brain for different target tasks.

It is likely that the ability of auditory analysis foundation models will continue to improve in the coming years, while adaptation and fine-tuning methods will likewise improve. Since time duration of sounds and events has been cited as a problematic issue for balancing feature size and resolution, it seems likely that advances will me made in the area of more effective multi-scale analysis methods.

\subsection{Future directions}

As we have explored, deep learning methods that make use of spectrogram features have gained prominence across the field of audio analysis, and for several speech analysis tasks too. However there are general aspects in which performance needs to improve substantially before widespread deployment becomes possible. These include the following;
\begin{itemize}
    \item Noise robustness, particularly towards overlapping sounds and reverberation.
    \item Model complexity and real-time operation on edge devices.
    \item Robust separation of intertwined sounds, particularly for polyphonic (multi recording channel) audio sources.
    \item Timeliness -- early detection before a sound has completed.
    \item Generalisation to unseen sounds, e.g. few- and zero-shot classification, including from multimodal prompts~\cite{PengfeiACMMM2025}.
\end{itemize}

Beyond this, determining optimal spectrogram settings for a particular back-end architecture and task is currently a largely empirical process. These settings refer to the dimension of spectrogram patches, their resolution in time and frequency domain, and whether the frequency dimension is linear or non-linear. Then each element (pixel) in a spectrogram must be scaled, such as log/A-law, perceptually scaled or otherwise. Data-driven methods of determining the optimal settings are required - but this conflicts with the aim of adopting generalised pre-trained foundation models, unless those can be multi-resolution and multi-scaled, or make use of data fusion techniques. We also note that many researchers rely on Python libraries such as librosa\footnote{\url{https://librosa.org}}, adopting default settings, so any future techniques should ideally be as stable and easy to use.

\authorcontributions{Authors contributed mainly to the sections relating to their respective domain expertise, with all authors contributing equally to the remaining sections of the manuscript. All authors have read and agreed to the published version of the manuscript.}

\funding{This research received no external funding.}


\conflictsofinterest{The authors declare no conflicts of interest.} 



\abbreviations{Abbreviations}{
The following abbreviations are used in this manuscript:
\\

\noindent 
\begin{tabular}{@{}ll}
AE & Autoencoder\\
AIM & Auditory Image Model\\
ASA & Acoustic Scene Analysis\\
ASD & Anomalous Sound Detection\\
AST & Audio Spectrogram Transformer\\
CNN & Convolutional Neural Network\\
CQT & Constant-Q Transform\\
DCASE & Detection and Classification of Acoustic Scenes and Events\\
DCT & Discrete Cosine Transform\\
DFT & Discrete Fourier Transform\\
DID & Dialect Identification\\
DWT & Discrete Wavelet Transform\\
ERB & Equivalent Rectangular Banks\\
FB & Filterbanks\\
FFT & Fast Fourier Transform\\
GTG & Gammatonegram\\
LID & Language Identification\\
LMS & Log-Mel Spectrogram\\
LNS & Log Non-uniform Spectrum\\
LS & Linear Spectrogram\\
LSS & Log-Scaled Spectrogram\\
LSTB & Long-Short Term Memory\\
MFCC & Mel-frequency Cepstral Coefficients\\
MS & Mel Spectrogram\\
OE & Outlier Exposure\\
PAMIR & Passive-aggressive Model for Image Retrieval\\
PLP & Perceptual Linear Prediction\\
PSDS & Polyphonic Sound Detection Score\\
RWCP & Real World Computing Partnership\\
RNN & Recurrent Neural Network\\
SAI & Stabilised Auditory Image\\
SDC & Shifted Delta Cepstra\\
SED & Sound Event Detection\\
SID & Speaker Identification\\
SSA & Sound Scene Analysis\\
SSD & Sound Scene Detection\\
SER & Speech Emotion Recognition\\
SNR & Signal to Noise Ratio\\
SV & Speaker Verification\\
SVM & Support Vector Machine\\
STFT & Short Time Fourier Transform\\
VAE & Variational Autoencoder\\
VNF & Variance Normalised Features
\end{tabular}
}

\begin{adjustwidth}{-\extralength}{0cm}

\reftitle{References}



\PublishersNote{}
\end{adjustwidth}
\end{document}